\documentclass[%
aps,
prl,
reprint,
superscriptaddress,
amsmath,
amssymb,
]{revtex4-2}

\usepackage{graphicx}
\usepackage{diagbox}
\usepackage{xcolor}
\usepackage{amsmath,amssymb,amsthm,amsfonts}
\usepackage{bbm}
\usepackage{bm}
\usepackage[normalem]{ulem}
\begin{document}

\title{Pure momentum-shift bulk photovoltaic effect in ferroelectric flat-band Mott insulators}

\author{Zhuocheng Lu}
\thanks{These authors contributed equally to this work.}
\affiliation{Center for Quantum Matter, School of Physics, Zhejiang University, Zhejiang 310058, China}

\author{Zhihao Gong}
\thanks{These authors contributed equally to this work.}
\affiliation{Academy of Interdisciplinary Studies on Intelligent Molecules, Tianjin Key Laboratory of Structure and Performance for Functional Molecules, College of Chemistry, Tianjin Normal University, Tianjin 300387, China}

\author{Jingshan Qi}
\affiliation{Tianjin Key Laboratory of Quantum Optics and Intelligent Photonics, School of Science, Tianjin University of Technology, Tianjin 300384, China}

\author{Hua Wang}
\email{daodaohw@zju.edu.cn}
\affiliation{Center for Quantum Matter, School of Physics, Zhejiang University, Zhejiang 310058, China}

\author{Kai Chang}
\email{kchang@zju.edu.cn}
\affiliation{Center for Quantum Matter, School of Physics, Zhejiang University, Zhejiang 310058, China}

\begin{abstract}
The shift current photovoltaic effect is conventionally understood as the real-space displacement of a wave packet induced by photoexcitation. However, this interpretation becomes insufficient in flat-band systems, where quasiparticles are too massive to accelerate in real space under the optical electric field. Here, we developed a physically consistent method to decompose the shift current into real-space and momentum-space components. A surprising pure momentum-space shift current is found theoretically in flat-band Mott insulator Nb$_3$X$_8$ (X = Cl, Br, I) monolayers. This work underscores that significant shift current responses can emerge even in systems with minimal interband polarization differences, highlighting the potential for exploring novel bulk photovoltaic effects in flat-band Mott insulators.
\end{abstract}

\maketitle

\textit{Introduction.}---Under monochromatic light irradiation, a direct current (DC) photocurrent, known as the bulk photovoltaic effect, arises in systems with broken spatial inversion symmetry. Initially, the bulk photovoltaic effect garnered attention for its potential to achieve high energy conversion efficiency beyond the Shockley-Queisser limit, which restricts the traditional photovoltaic effect \cite{spanier2016, dai2023}. More recently, this effect has become a focus of research due to its connection to fundamental properties of quantum materials, including symmetry, order parameters (such as ferroelectricity and ferromagnetism), topology, and many-body interactions \cite{hosur2011, young2012, morimoto2016, nakamura2017, chan2017, ma2019, zhang2019, gao2020, wang2020, rees2020, dejuan2017, ahn2020, chaudhary2022}. This connection holds promise for applications in material characterization and the development of devices based on novel quantum transport and optical phenomena. 

In intrinsic semiconductors, when photon energy exceeds the bandgap, shift current becomes one of the primary mechanisms driving the bulk photovoltaic effect \cite{sipe2000, kaplan2022}. Shift current refers to the photocurrent generated by the displacement between the centers of electron wavefunctions in the conduction and valence bands during photoexcitation, traditionally interpreted as a real-space phenomenon. Based on recent studies employing an anomalous acceleration framework, it has been shown that the shift current arises from contributions in both real and momentum spaces \cite{holder2020, kaplan2022}. Within this framework, for dispersive-band systems, the extended nature of the wavefunctions combined with their small effective masses leads to a significant real-space displacement upon photoexcitation. In contrast, in flat-band systems, where wavefunctions are highly localized in real space, the framework indicates that real-space displacement is minimal, making momentum-space displacement the dominant contributor.

However, the previous treatment of the real-space shift as the difference in the gauge-dependent Berry connections between the conduction and valence bands results in an ambiguous interpretation of both real-space and momentum-space shifts, and limits discussions to model Hamiltonians \cite{kaplan2022}. A prototypical topological flat-band system, e.g. twisted bilayer graphene \cite{cao2018a, cao2018, lu2019, yankowitz2019, cao2020}, has been considered an ideal candidate for realizing such photovoltaic effect \cite{kaplan2022}. Nonetheless, its experimental realization is hindered by graphene's zero bandgap and precise control required for twisted angles.


To address these challenges, this work aims to develop a physically consistent theory to decompose the real-space and momentum-space contributions to the shift current, enabling first-principles studies of this phenomenon in flat-band systems. Moreover, we focus on Nb$_3$X$_8$ monolayers, which provide an ideal platform for investigating the impact of correlations and topological flat bands on nonlinear optical responses \cite{sun2022, regmi2022, gao2023, feng2023, grytsiuk2024}. In these systems, the topological flat bands arise from the destructive interference of wavefunctions in a breathing Kagome lattice. Recent studies have further identified Nb$_3$X$_8$ monolayers as long-sought single-band Mott insulators, simplifying the low-energy physics to intrinsically correlated systems\cite{lee2006, gao2023, grytsiuk2024}.

In this work, we develop a comprehensive theory about pure momentum-shift current in flat-band Mott insulator to decompose the shift current into real-space and momentum-space contributions, establishing a physically consistent approach to understanding the bulk photovoltaic effect in topological flat-band systems. We begin by introducing the microscopic expression for shift current, which is proportional to the sum of the products of the shift vector and the absorption strength across all $\boldsymbol{k}$-points. We propose that the real-space shift is described by the difference between the Wannier centers of the conduction and valence bands. By subtracting the real-space shift from the shift vector, we obtain the momentum-space shift. This decomposition allows us to isolate the real-space and momentum-space contributions to the shift current during photoexcitation. With this framework, we perform first-principles calculations on $\rm Nb_3X_8$ monolayers and analyze the characteristics of their shift current. Our results reveal that their maximum peaks in shift current spectra are entirely driven by the momentum-space shift, with minimal real-space contribution. Additionally, we examine the relationship between the shift current, collective shift vector and the imaginary part of the dielectric function, finding that their magnitudes are generally correlated. 


\textit{Comprehensive theory of shift current.}---For a system illuminated by linearly polarized light $E^b(t) = E^b(\omega)e^{i\omega t} + E^b(-\omega)e^{-i\omega t}$, the shift current can be formulated as
\begin{equation}
J_{\text{SC}}^a = \sigma_{\text{SC}}^{abb} E^b(\omega) E^b(-\omega),
\label{eq:II-1}
\end{equation}
where $J_{\text{SC}}^a$ is the current density arising from the shift mechanism, and $\sigma_{\text{SC}}^{abb}=\sigma_{\text{SC}}^{abb}(0;\omega,-\omega)$ is the shift current photoconductivity, with the frequency dependence omitted here for simplicity. According to nonlinear optical theory, this photoconductivity can be expressed as \cite{aversa1995,sipe2000,wang2020}:
\begin{equation}
\sigma_{\text{SC}}^{abb} = -e \int_{\boldsymbol{k}} \sum_{n, m} S_{mn}^{ab} \epsilon_{nm}^{bb}(\omega),
\label{eq:II-2}
\end{equation}
where $e = |e|$ is the electron charge, $\int_{\boldsymbol{k}} = \int d^d \boldsymbol{k}/(2\pi)^d$, $S_{mn}^{ab}$ is the well-known shift vector describing the extent of wavepacket displacement, and $\epsilon_{nm}^{bb}(\omega)$ characterizes the absorption strength during photoexcitation. Specifically, $S_{mn}^{ab} = r_{mm}^a - r_{nn}^a - \partial_{k^a} \varphi_{mn}^b$, where $r_{nn}^a = \langle u_n | i \partial_{k^a} | u_n \rangle$ is the intraband Berry connection, and $\varphi_{mn}^b$ is the phase of the interband Berry connection $r_{mn}^b = |r_{mn}^b| e^{i \varphi_{mn}^b}$. Additionally, $\epsilon_{nm}^{bb}(\omega) = f_{nm} \delta(\omega_{mn} - \omega) \Gamma_{nm}^{bb}$, where $f_{nm} \delta(\omega_{mn} - \omega)$ denotes the joint density of states, and $\Gamma_{nm}^{bb}=\pi e^2/\hbar^2 |r_{mn}^b|^2 $ represents the transition probability. Here, $f$ is the Fermi-Dirac distribution with $f_{nm} \equiv f_n - f_m$, and $\delta(\omega_{mn} - \omega)$ is the Dirac delta function enforcing energy conservation during photoexcitation.

\begin{figure}[htbp]
    \centering
    \includegraphics[width=\columnwidth]{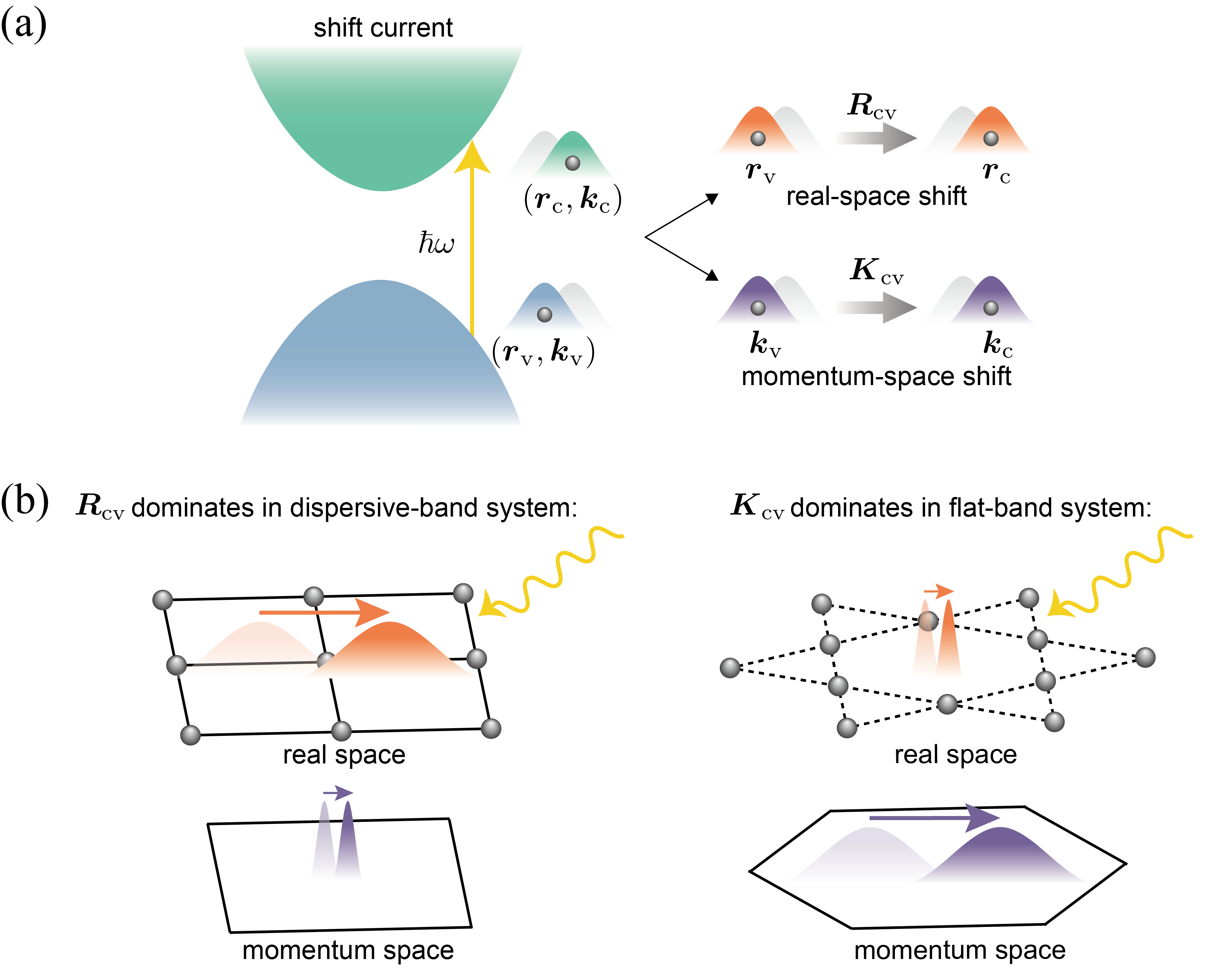}
    \caption{Shift current mechanism (a) and two typical conditions under which the real-space shift or momentum-space shift dominates (b). $(\boldsymbol{r}_{\rm v},\boldsymbol{k}_{\rm v})$ and $(\boldsymbol{r}_{\rm c},\boldsymbol{k}_{\rm c})$ denote the centers of wavefunctions for electrons in the valence and conduction bands, respectively.}
    \label{f01}
\end{figure}

Building on this understanding, we aim to decompose the shift current into real-space and momentum-space components in a physically meaningful way. First, we recognize that the wavepacket shift in real space during photoexcitation is described by the change in the Wannier center, $R_{mn}^a = R_m^a - R_n^a$, where $R_n^a = \langle w_{n\boldsymbol{0}} | r^a | w_{n\boldsymbol{0}} \rangle$. Next, we obtain the wavepacket’s shift in momentum space by subtracting the shift vector $S_{mn}^{ab}$ from the real-space shift $R_{mn}^a$:
\begin{equation}
K_{mn}^{ab} = S_{mn}^{ab} - R_{mn}^a.
\label{eq:II-3}
\end{equation}
This decomposition remains invariant provided a specific unit cell is chosen to resolve the ambiguity in the Wannier center difference, $R_{mn}^a$. As illustrated in Fig. \ref{f01}(a), the shift current photoconductivity can be decomposed into two components, $\sigma_{\text{SC}}^{abb} = \sigma_{\text{R}}^{abb} + \sigma_{\text{K}}^{abb}$, where $\sigma_{\text{R}}^{abb}$ and $\sigma_{\text{K}}^{abb}$ represent contributions from the real-space shift ($R_{mn}^a$) and momentum-space shift ($K_{mn}^{ab}$), respectively. The decomposition of shift current in dispersive-band and flat-band systems is schematically illustrated in Fig. \ref{f01}(b) \footnote{We observe that a divergent shift vector and zero absorption result in a finite shift current in the SnTe monolayer [npj Comput. Mater. \textbf{10}, 23 (2024)]. Therefore, while optical transitions induce a large real-space shift due to the dispersive nature of the bands, the momentum-space shift current dominates. However, in most cases, the bulk photovoltaic effect is correlated with a finite optical absorption process. We will restrict our discussion to this scenario.}.

According to Eq. (\ref{eq:II-3}), we can obtain the decomposition of the shift current. However, the interpretation of the momentum-space shift remains unclear. To address this issue, we consider the integration of photoconductivity over frequency, a quantity more relevant to real-world applications \cite{rangel2017}, and directly related to the integral of the shift vector, which can be expressed as follows:
\begin{equation}
\int_{\omega} \sigma^{a b b}_{\rm SC} \approx \Gamma^{bb} \sum_{n, m} f_{n m}\left(P_{mn}^a -e W_{mn}^{ab}\right),
\label{eq:II-4}
\end{equation}
where transition probability is approximated by an averaged value $\Gamma^{bb}_{m n} \approx \Gamma^{bb}$, $P_{mn}^a=P_m^a-P_n^a$ is the interband polarization difference with $P_n = -e R_n^a$ according to the modern theory of polarization \cite{king-smith1993}, $W_{mn}^{ab}=\int_{\boldsymbol{k}}\partial_{k^a}\varphi_{mn}^b$ is the winding of the phase \cite{fregoso2017}. From our previous discussion, it is evident that $P_{mn}^a$ and $W_{mn}^{ab}$ correspond to the real-space and momentum-space shifts, respectively. If optical zeros ($r_{mn}^b=0$ at certain $\boldsymbol{k}$ points in the Brillouin zone) are absent, the total response will arise solely from the real-space shift current, as the momentum-space shift current can be eliminated by explicitly choosing the optical gauge where $\varphi_{m n}^b=0$. However, optical zeros are commonly present in real materials, leading to a discontinuity in $\partial_{\boldsymbol{k}}\varphi_{mn}^b$ and causing $W_{mn}^{ab}$ to affect the total response. As a result, the momentum-space shift current can be attributed to these singular optical zeros in the Brillouin zone, which can significantly contribute to the shift current. 

Before concluding this section, we want to discuss the feasibility of distinguishing between the real-space shift current and the momentum-space shift current. We note that for a system describable within the two-band approximation, $\sigma_{\text{R}}^{abb}$ simplifies to
\begin{equation}
\sigma_{\text{R}}^{abb} = -\frac{e}{\hbar} \Delta R^a \epsilon^{bb}(\omega),
\label{eq:II-5}
\end{equation}
where $\Delta R^a = R_{\rm cv}^a$, with $\rm c$ and $\rm v$ denoting the conduction and valence bands, respectively. Here, $\epsilon^{bb}(\omega) = \int [d\boldsymbol{k}] \sum_{n, m} \epsilon_{nm}^{bb}(\omega)$ is the imaginary part of the dielectric function \cite{sipe2000, ibanez2018}. The change in the Wannier center, $\Delta R^a$, is factored out of the integral since it is independent of $\boldsymbol{k}$, and the summation over band indices is omitted in the two-band case. In Eq. (\ref{eq:II-5}), $\sigma_{\text{R}}^{abb}$ has been expressed as the product of two measurable quantities. This implies that our discussion of the real-space and momentum-space shift currents could potentially be verified.



\begin{figure}[htbp]
    \centering
    \includegraphics[width=\linewidth]{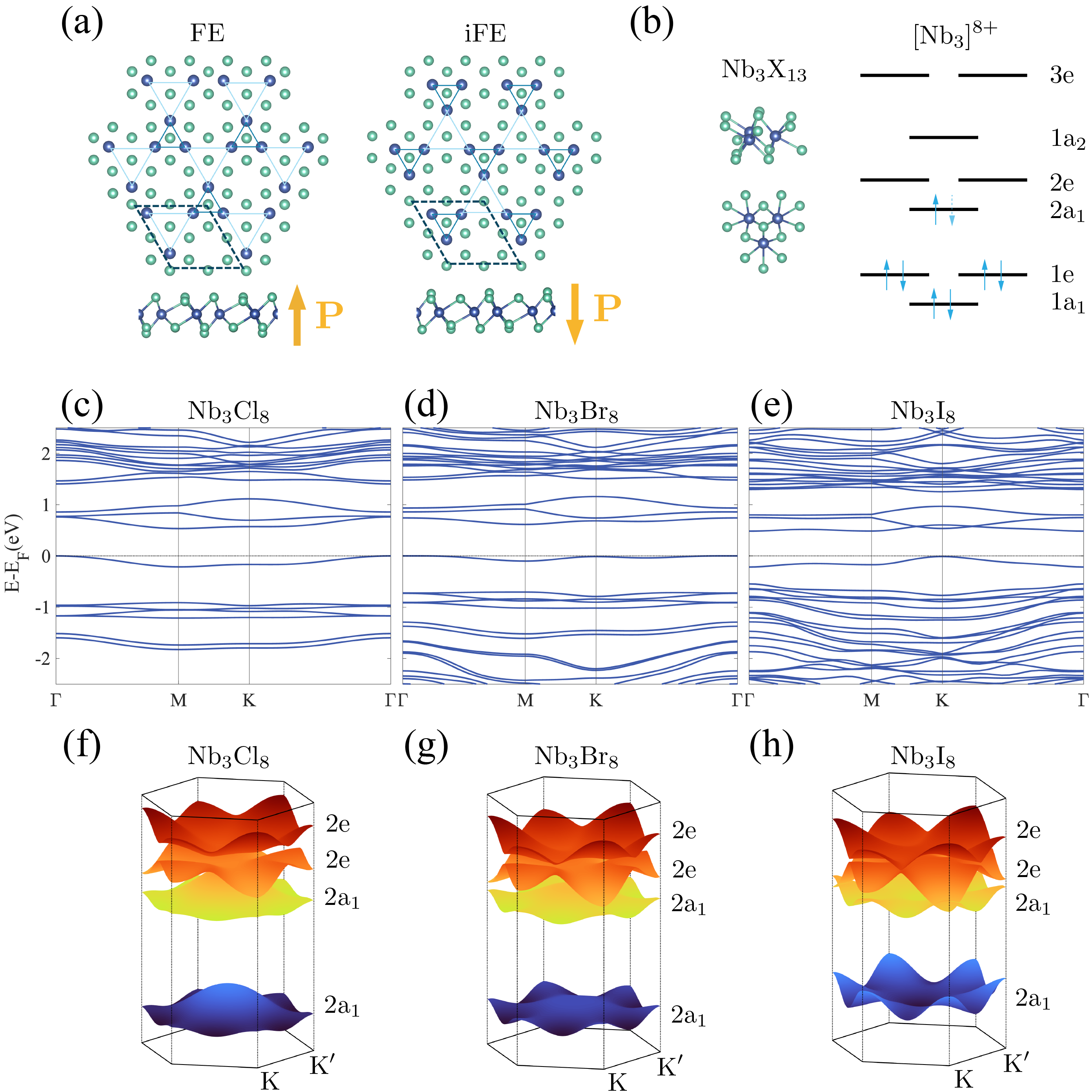}
    \caption{Crystal structure and band structure of $\rm Nb_{3}X_{8}$ monolayers. (a) Two different ferroelectric configurations. Blue atoms indicate Nb, while green atoms represent X (X = Cl, Br, I). Orange arrows indicate the polarization. Dark (light) blue lines connect Nb atoms within the same (different) trimers. The black dashed lines indicates the primitive cell. (b) $\rm Nb_{3}X_{13}$ cluster and schematic picture of $\rm [Nb_{3}]^{8+}$ molecular orbitals. Solid (dashed) blue arrows denote occupied and unoccupied states, respectively. (c-e) Band structures of FE configurations along high-symmetry lines. (f-h) 2D band plot of four isolated bands near the Fermi levels. The correspondence between these bands and the $\rm [Nb_{3}]^{8+}$ molecular orbitals is depicted on the right side. }
    \label{f02}
\end{figure}

\textit{Shift current in $Nb_3X_8 (X = Cl, Br, I)$ monolayers.}---To validate our shift current theory, we performed first-principle calculations of shift current and its decomposition in the topological flat-band system $\rm Nb_3X_8$ monolayers. $\rm Nb_3X_8$ monolayers belong to space group P3m1. The ferroelectricity in these systems arise from inversion symmetry breaking induced by Nb trimerization. Two ferroelectric configurations with opposite out-of-plane polarizations are depicted in Fig. \ref{f02}(a). As schematically illustrated in Fig. \ref{f02}(b), the electrons in $\rm Nb_3X_8$ monolayers resemble the molecular orbitals of Nb-Nb metal bonds within a $\rm Nb_3X_{13}$ cluster, as the $\rm Nb_3X_8$ monolayers can be considered to form through the interconnection of $\rm Nb_3X_{13}$ clusters \cite{cotton1964, bursten1982, kennedy1996, sheckelton2017, haraguchi2017, gao2023}. For concreteness, we focus on the FE configurations and assume that the magnetic moment of the Nb trimer is oriented along the positive $z$ direction. This assumption is reasonable because the shift current is even under time-reversal symmetry and is thus insensitive to the direction of the magnetic moment. The details of our first-principles calculations for $\rm Nb_3X_8$ monolayers are provided in Supplementary. As shown in Fig. \ref{f02}(c-e), the low-energy physics of these monolayers is dominated by four isolated bands directly associated with the $\rm [Nb_{3}]^{8+}$ molecular orbitals. The 2D plots of these four bands are presented in Fig. \ref{f02}(f-h).

\begin{figure}[htbp]
    \centering
    \includegraphics[width=\linewidth]{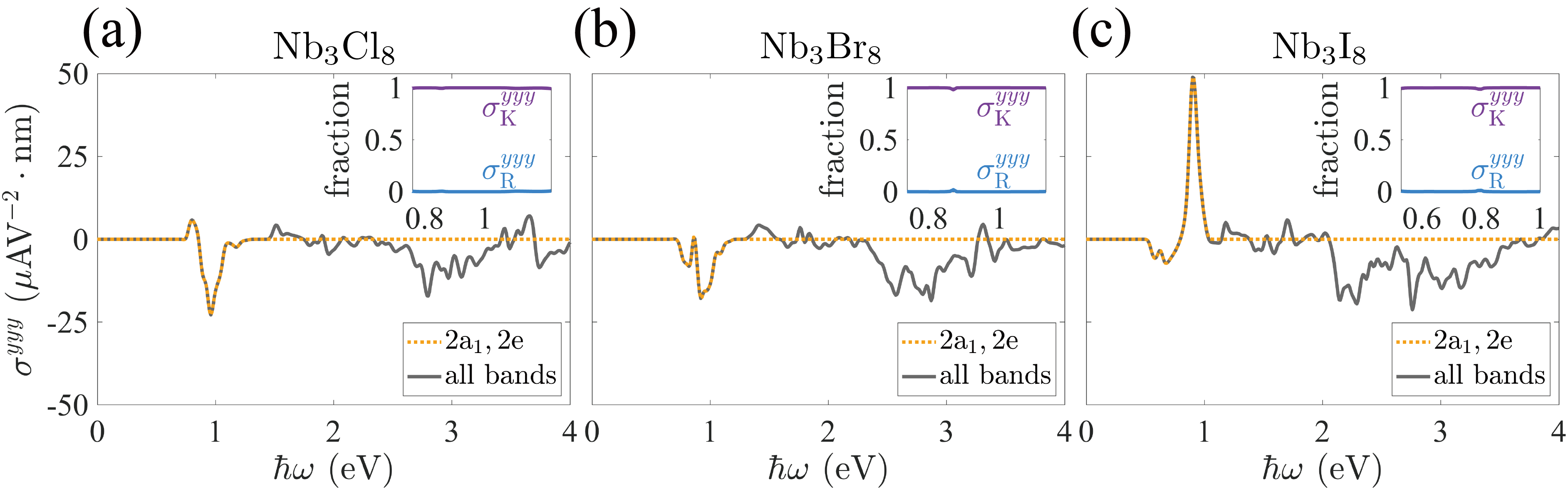}
    \caption{Shift current spectra of $\rm Nb_{3}X_{8}$ monolayers (a-c). Dark grey lines indicate the total response, while orange lines are the contribution arising from the occupied 2a$_1$ orbital to the unoccupied 2a$_1$ and 2e orbitals. The insets at the top right show the fractions of real-space shift and momentum-space shift. The fraction is defined as $n_{\rm R/K}=|\sigma^{yyy}_{\rm R/K}|/(|\sigma^{yyy}_{\rm R}|+|\sigma^{yyy}_{\rm K}|)$. Blue lines represent real-space shifts, and purple lines represent momentum-space shifts. }
    \label{f03}
\end{figure}

Shift current spectra of Nb$_3$X$_8$ monolayers are calculated and plotted as the gray lines in Fig. \ref{f03}. Here, we only present the results of $\sigma^{yyy}$ component. Because the shift current photoconductivity of Nb$_3$X$_8$ monolayers has only one independent non-zero component according to symmetry analysis, specifically $\sigma^{yyy} = -\sigma^{yxx} = -\sigma^{xxy} = -\sigma^{xyx}$. Based on the discussion in the previous section, we further calculated the contributions of the real-space shift and momentum-space shift to the total response and plotted their fractions in the inset of Fig. \ref{f03}. We were surprised to find that the maximum peaks of shift current spectra in Nb$_3$X$_8$ monolayers originate entirely from the momentum-space shift. To illustrate its origin, we separately calculated the contributions of excitations from the occupied 2a$_1$ molecular orbital to the unoccupied 2a$_1$ and 2e molecular orbitals for the shift current, and plotted these contributions as orange lines in Fig. \ref{f03}. We observed that this contribution completely overlaps with the total response at these peaks, indicating that these molecular orbitals are responsible for both the peaks and the pure momentum-space shift. Further symmetry analysis revealed it is the C$_{\rm 3z}$ symmetry that enforces these molecular orbitals to be located at the center of the Nb trimer, which results in a zero real-space shift during photoexcitation. Therefore, this pure momentum-space shift current is robust to deformation and disorder, as long as the C$_{\rm 3z}$ symmetry is preserved. Despite we did not present the decomposition of the shift current at higher energies, it can be anticipated that as more bands participate in photoexcitation, the real-space shift current will become non-zero and may even dominate.

\begin{figure}[htbp]
    \centering
    \includegraphics[width=\linewidth]{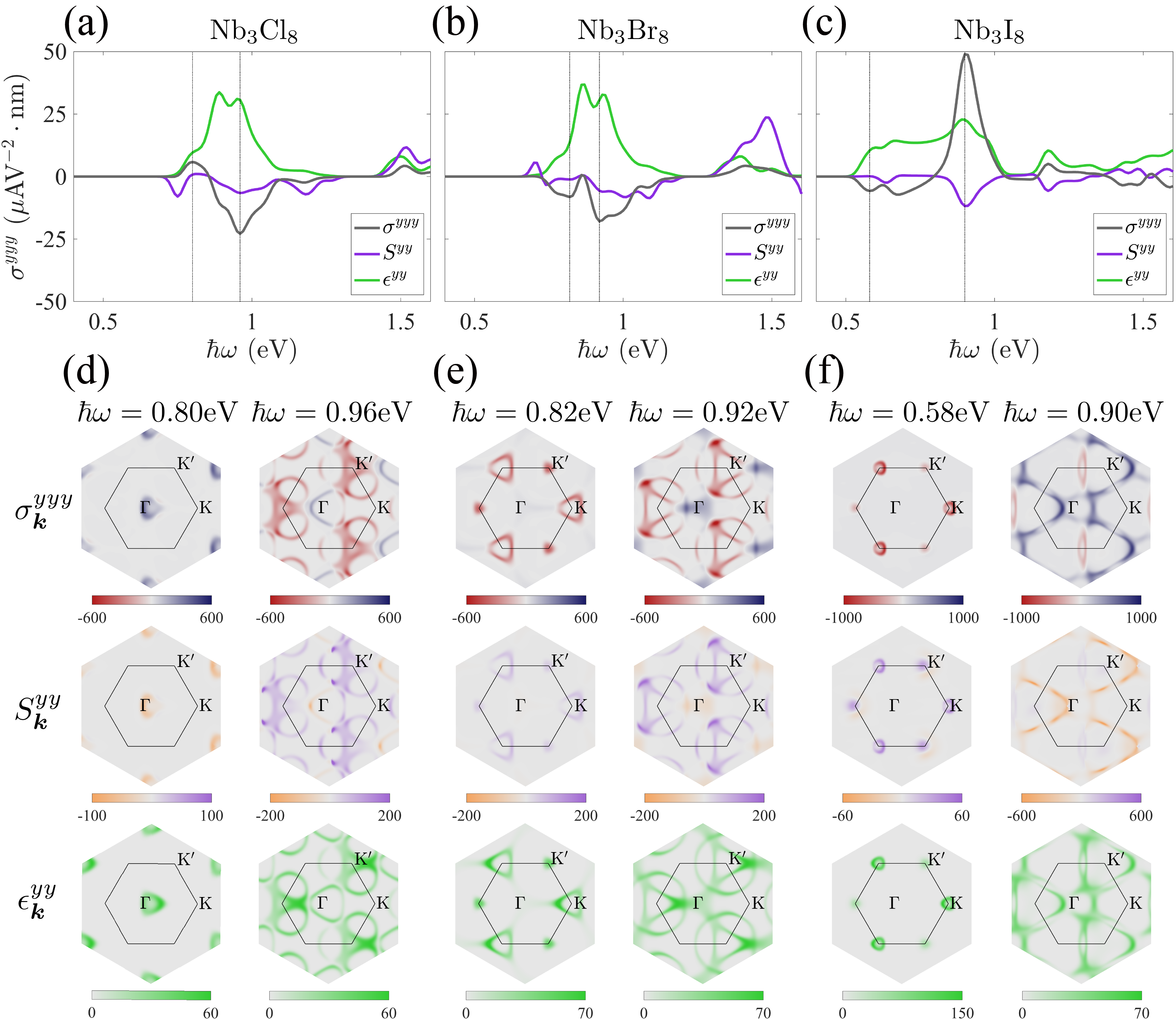}
    \caption{Collective shift vector $S^{yy}$ and imaginary part of the dielectric function $\epsilon^{yy}$ compared with shift current photoconductivity $\sigma^{yyy}$ in the low energy region (a-c) and $\boldsymbol{k}$ space distribution of these quantities in $\rm Nb_{3}X_{8}$ monolayers (d-f). The dark gray lines mark $\sigma^{yyy}$, the purple lines represent $S^{yyy}$, and the green lines indicate $\epsilon^{yy}$. The $\boldsymbol{k}$-resolved shift current photoconductivity, collective shift vector, and imaginary part of the dielectric function are marked with $\sigma^{yyy}_{\boldsymbol{k}}$, $S^{yy}_{\boldsymbol{k}}$ and $\epsilon^{yy}_{\boldsymbol{k}}$, respectively. }
    \label{f04}
\end{figure}

To clarify the influence of the shift vector and the imaginary part of the dielectric function on the magnitude of the shift current photoconductivity, we further studied the relationship between the spectra of $\sigma^{yyy}$, $\epsilon^{yy}$ and $S^{yy}$ and plotted them in Fig. \ref{f04}(a-c), where the collective shift vector $S^{yy}$, defined as
\begin{equation}
    S^{yy} = \int_{\boldsymbol{k}} \sum_{n, m} f_{n m} S_{m n}^{y y} \delta\left(\omega_{m n} - \omega\right),
\label{eq:III-1}
\end{equation}
characterizes the displacement strength of wavepacket centers in real space and momentum space during photoexcitation. Since $S^{yy}$ and $\epsilon^{yy}$ have different units, they are scaled by a constant factor for easier comparison. While the maximum peak of $\epsilon^{yy}$ in the $\rm Nb_{3}I_{8}$ monolayer is slightly lower than that in $\rm Nb_{3}Cl_{8}$ and $\rm Nb_{3}Br_{8}$ monolayers, the notably larger $S^{yy}$ in $\rm Nb_{3}I_{8}$ results in the highest $\sigma^{yy}$ among the $\rm Nb_{3}X_{8}$ monolayers. This indicates that the magnitude of $\sigma^{yyy}$ is roughly proportional to the product of $S^{yy}$ and $\epsilon^{yy}$. In Fig. \ref{f04}(d-f), we show the $\boldsymbol{k}$-resolved $\sigma^{yyy}$, $S^{yy}$ and $\epsilon^{yy}$ corresponding to the first and maximum peaks in the shift current spectra of $\rm Nb_{3}X_{8}$ monolayers. We can see that the first peaks arise from photoexcitations near the band edges, while the maximum peaks involve contributions from photoexcitations distributed across the entire Brillouin zone. Therefore, the maximum peaks in the shift current spectra can be attributed to the enhancement of large joint density of states in flat-band systems. 

\textit{Discussion and conclusion.}---In summary, we propose a physically consistent theory that decomposes the shift current into two contributions: real-space and momentum-space shifts. Using this theory, we investigated the shift current in the flat-band systems $\rm Nb_{3}X_{8}$ monolayers. We found that the maximum peak of the shift current spectra is driven by photoexcitation between the molecular orbitals of the Nb trimers, originating entirely from the momentum-space shift contribution. This pure momentum-space shift current remains robust as long as the C$_{\rm 3z}$ symmetry is preserved. Additionally, we discussed the connections between the shift current photoconductivity, collective shift vector, and imaginary part of the dielectric function. As detailed in Supplementary, we also studied the injection current spectra of $\rm Nb_{3}X_{8}$ monolayers. We found that the magnitude of the injection current is negligible compared to that of the shift current. This characteristic can be attributed to the weakly broken time-reversal symmetry and the flatness of the bands in $\rm Nb_{3}X_{8}$ monolayers. Our research provides a crucial distinction between the bulk photovoltaic effects observed in topological flat-band systems and conventional systems, expanding the search for materials that exhibit a large bulk photovoltaic effect.

\textit{Acknowledgments.}---H.W. acknowledges the support from the NSFC under Grants Nos. 12304049 and 12474240, as well as the support provided by the Zhejiang Provincial Natural Science Foundation of China under grant number LDT23F04014F01. K. C. acknowledges the support from the Strategic Priority Research Program of the Chinese Academy of Sciences (Grants Nos. XDB28000000 and XDB0460000), the NSFC under Grants Nos. 92265203 and 12488101, and the Innovation Program for Quantum Science and Technology under Grant No. 2024ZD0300104.


\clearpage
\appendix
\onecolumngrid

\section*{Supplementary}
\subsection{A. Numerical calculation details}
To access the electronic structures of Nb$_{3}$X$_{8}$ monolayers, first-principles calculations were performed using the Vienna ab initio Simulation Package (VASP) equipped with the projector-augmented-wave potentials \cite{kresse1996}. The exchange-correlation interactions were considered in the generalized gradient approximation (GGA) with Perdew-Burke-Ernzerhof scheme \cite{perdew1996}. A 400 eV cutoff energy and $6\times6\times1$ $\boldsymbol{k}$-point sampling were considered for relaxation and self-consistent calculations. The structures were optimized until the force on each atom is reduced to below 0.01 eV/Å. The correlation of $4d$ orbitals of $\text{Nb}$ was treated with GGA+$U$ method, with an isotropic $U=2.0$ eV \cite{jiang2017, regmi2022, hu2023}. Spin-orbit coupling (SOC) was included in all calculations. To calculate the photoconductivities of the shift current and injection current, as well as the $\boldsymbol{k}$-resolved quantities, we constructed a Wannier tight-binding model using the \texttt{wannier90} package and utilized a modified \texttt{postw90} program \cite{pizzi2020}. To calculate the injection current conductivity, we used a moderate relaxation time of $\tau = 1.0 \times 10^{-13}$ s.

\subsection{B. Effects of Hubbard $U$}
\begin{figure}[htbp]
    \centering
    \includegraphics[width=0.6\linewidth]{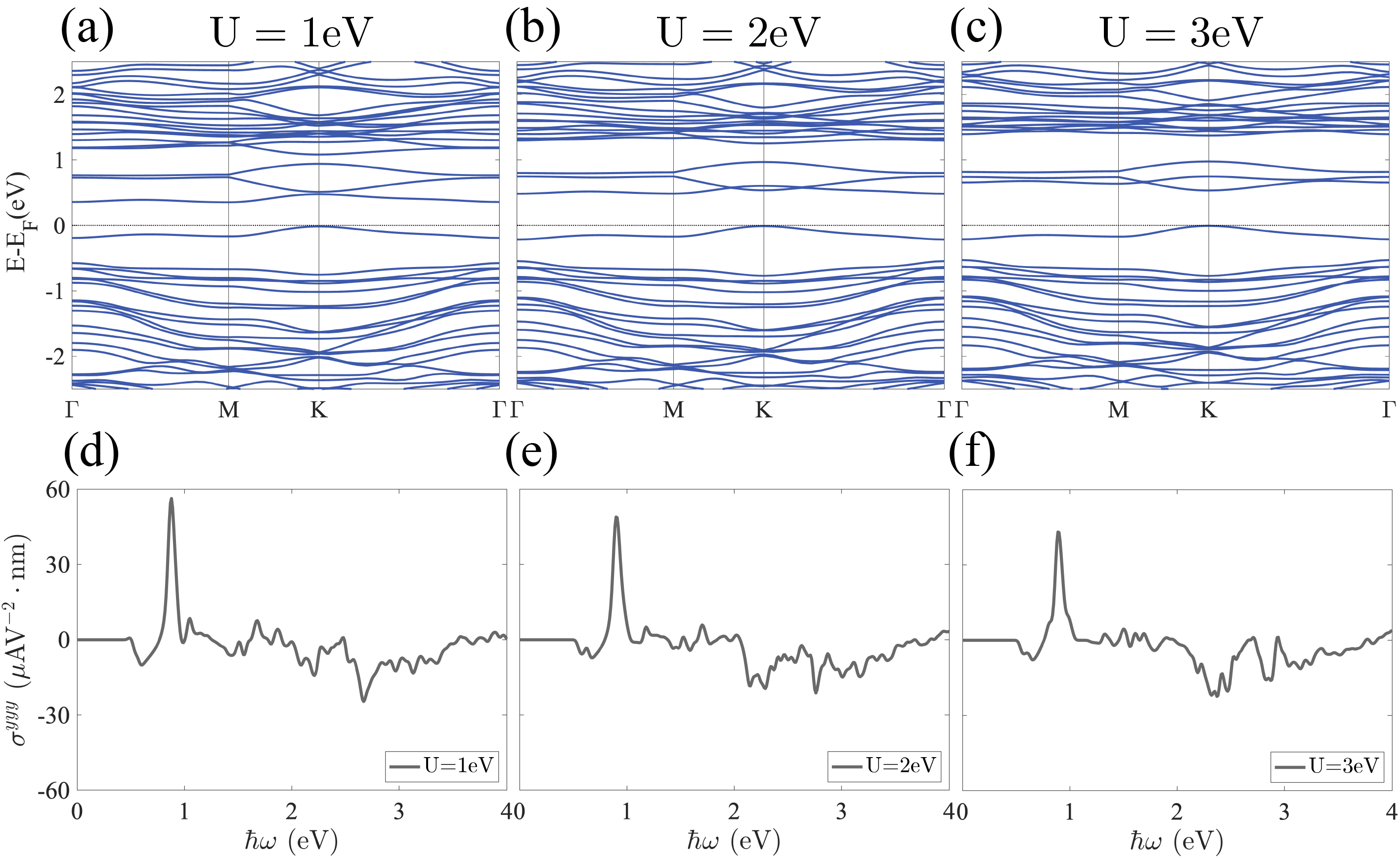}
    \caption{Band structures (a-c) and shift current spectra (d-f) in $\rm Nb_{3}I_{8}$ monolayers with Hubbard $U$=1,2, and 3 eV, respectively.}
    \label{f05}
\end{figure}

\renewcommand\theequation{B\arabic{equation}}
\setcounter{equation}{0} 
As discussed in the Section A of Supplementary, we used a widely adopted Hubbard $U = 2$ eV to describe the correlation effects in $\rm Nb_{3}X_{8}$ monolayers \cite{jiang2017, regmi2022, hu2023}. However, since the strength of Hubbard $U$ can significantly affect the electronic structure and transport properties in correlated systems, it is crucial to access the impact of the choice of $U$. In this section, we perform first-principles calculations for the $\rm Nb_{3}I_{8}$ monolayer, which exhibits the highest response among the $\rm Nb_{3}X_{8}$ monolayers, using Hubbard $U = 1$, 2, and 3 eV to investigate its influence on the numerical results. In Fig. \ref{f05}(a-c), we show the band structures for Hubbard $U = 1$, 2, and 3 eV, respectively. As shown, four bands near the Fermi level remain distinct and are separated from other bands. Additionally, as $U$ increases from 1 eV to 3 eV, the overlap between the unoccupied 2$a_1$ and 2$e$ orbitals becomes stronger. In Fig. \ref{f05}(d-f), we present the shift current spectra for $\rm Nb_{3}I_{8}$ monolayer with Hubbard $U = 1$, 2, and 3 eV, respectively. While the maximum peak slightly decreases as $U$ increases, the shift current spectra for different values of $U$ remain similar, with the peak in the low energy region consistently being the highest. Therefore, the conclusions in the main text are robust to variations in Hubbard $U$ across this range.

\subsection{C. Injection current in $\rm \mathbf{Nb_3X_8 (X = Cl, Br, I)}$ monolayers}
\renewcommand\theequation{C\arabic{equation}}
\setcounter{equation}{0} 

\begin{figure}[htbp]
    \centering
    \includegraphics[width=0.6\linewidth]{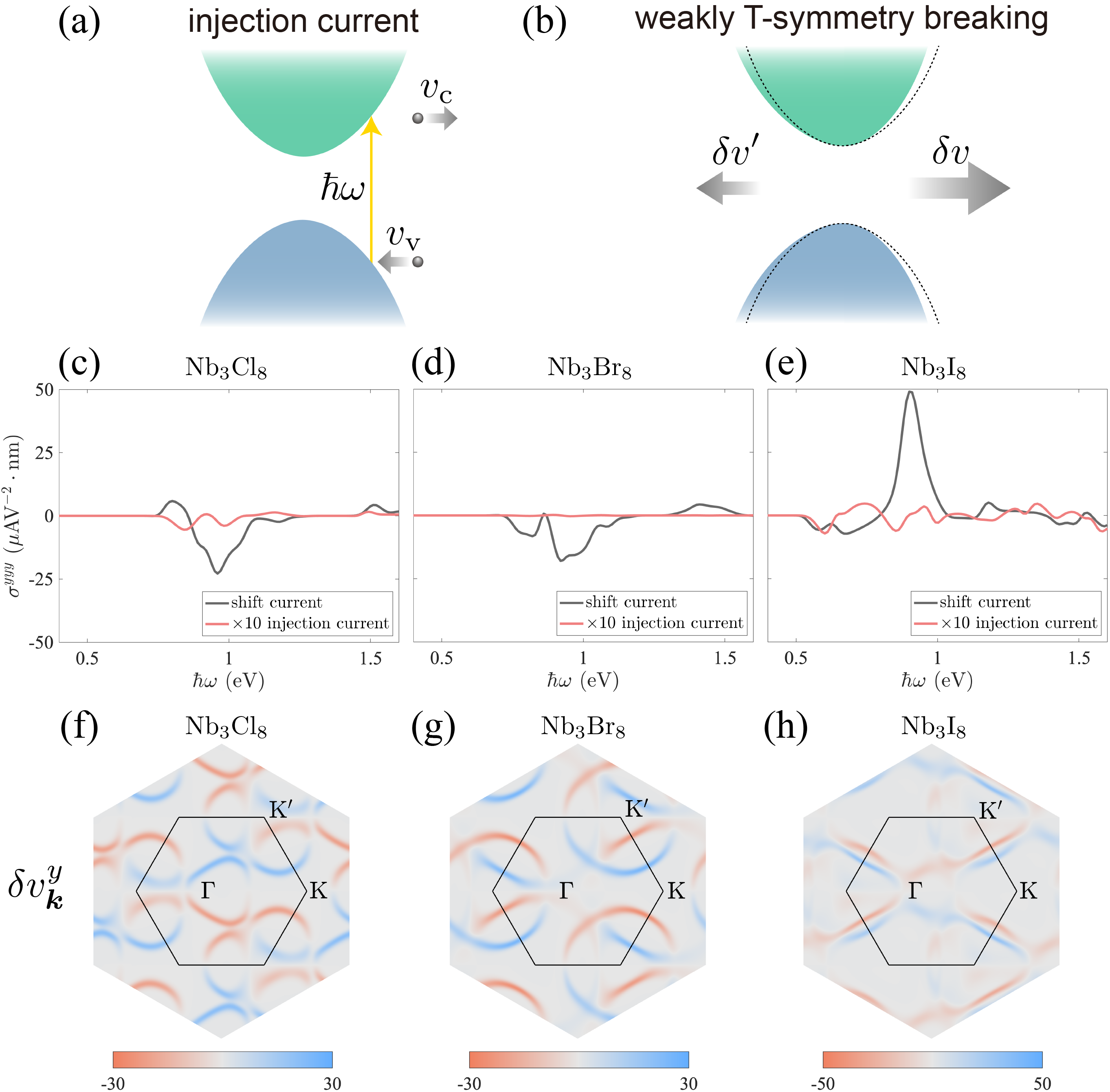}
    \caption{Schematic of injection current mechanism and injection current spectra in $\rm Nb_{3}X_{8}$ monolayers. (a) Schematic of injection current mechanism. Injection current arise from velocity difference between conduction band and valance band during photoexcitation. (b) Schematic of injection current in the system with weakly time-reversal broken. (c-e) Injection current spectra in $\rm Nb_{3}X_{8}$ monolayers. The injection current is indicated by light red lines, scaled up by a factor of ten for comparison. (f-h) $\boldsymbol{k}$-resolved collective velocity difference $\delta v^y_{\boldsymbol{k}}$ in $\rm Nb_{3}X_{8}$ monolayers.}
    \label{f06}
\end{figure}

Besides the shift current, another primary mechanism contributing to the bulk photovoltaic effect is the injection current \cite{sipe2000}, with a schematic illustration of the injection mechanism shown in Fig. \ref{f06}(a). The term "injection current" originates from the fact that the magnitude of this photocurrent is proportional to the illumination time when scattering mechanisms are neglected.  Microscopically, injection current can be interpreted as the photocurrent generated by the difference in group velocities of electrons in the conduction and valence bands during photoexcitation. Driven by linearly polarized light, the microscopic expression for the injection current photoconductivity is given by:
\begin{equation}
\sigma_{\text{IC}}^{abb} = -\frac{\tau \pi e^3}{\hbar^2 } \int_{\boldsymbol{k}} \sum_{n, m} \Delta_{mn}^a \epsilon_{nm}^{b b},
\label{eq:C-1}
\end{equation}
where $\tau$ is the relaxation time, and $\Delta_{mn}^a=v_{mm}^a - v_{nn}^a$ is the difference of group velocity between band $m$ and band $n$ along the $a$-direction. Other quantities are defined in the discussion of shift current photoconductivity in Section II. In contrast to $\sigma_{\text{SC}}^{abb}$, which is even under time-reversal symmetry, $\sigma_{\text{IC}}^{abb}$ is odd under time-reversal symmetry. Thus, this contribution only exists in magnetic systems where time-reversal symmetry is broken.

We calculated the injection current spectra in $\rm Nb_{3}X_{8}$ monolayers, as shown in Fig. \ref{f06}(c-e). The injection current photoconductivity remains significantly smaller than that of the shift current, even when scaled by a factor of ten. This characteristic of the bulk photovoltaic effect in $\rm Nb_{3}X_{8}$ monolayers can be explained as follows. First, as shown in the schematic in Fig. \ref{f06}(b), in systems with slightly broken time-reversal symmetry, contributions to the injection current from pairs of $\boldsymbol{k}$-points related by time-reversal symmetry tend to cancel each other. Thus, only systems with sufficiently strong magnetism can exhibit a significant injection current. In $\rm Nb_{3}X_{8}$ monolayers, the magnetic moment of 1 $\mu_B$ per Nb trimer induces only a relatively weak breaking of time-reversal symmetry. Therefore, the injection current in $\rm Nb_{3}X_{8}$ monolayers is expected to be correspondingly small. Additionally, the bands near the Fermi level in $\rm Nb_{3}X_{8}$ monolayers show minimal energy dispersion, leading to a small difference in group velocities between the conduction and valence bands during photoexcitation. This can help explain why the $\rm Nb_{3}Br_{8}$ monolayer, which has the narrowest bandwidth among the $\rm Nb_{3}X_{8}$ monolayers, exhibits the smallest injection current within this family. We attribute the significantly smaller injection current relative to the shift current in $\rm Nb_{3}X_{8}$ monolayers to the factors discussed above. To provide further insight into the extremely weak injection current in $\rm Nb_{3}X_{8}$ monolayers, we define a collective velocity difference, analogous to the collective shift vector, as follows:
\begin{equation}
\delta v^y= \int_{\boldsymbol{k}} \sum_{n, m} f_{n m}\Delta_{m n}^y \delta\left(\omega_{m n}-\omega\right),
\label{eq:C-2}
\end{equation}
which characterizes the magnitude of the velocity difference during photoexcitation. The distribution of $\delta v^y_{\boldsymbol{k}}$ across the Brillouin zone is shown in Fig. \ref{f06}(f-h). As expected, contributions in the Brillouin zone exhibit significant cancellation.


\end{document}